\documentclass[aps,pre,preprint,superscriptaddress,groupedaddress]{revtex4}     
\usepackage{latexsym}
\usepackage[T1]{fontenc}           
\usepackage{natbib}
\usepackage{amsmath}
\usepackage{amsfonts}
\usepackage{amssymb}
\usepackage[dvips]{color,graphicx}   
\usepackage{verbatim}   
\usepackage{t1enc}
\usepackage{anysize}
\usepackage{float}  
\usepackage{natbib}
\usepackage{array}
\usepackage[percent]{overpic} 
\usepackage{graphicx}
\usepackage{color}
\usepackage{lmodern}
\DeclareGraphicsExtensions{.pdf,.png,.jpg}

\begin{document}                  

\widetext

\title{Predicting the extinction of Ebola spreading in Liberia\\
due to mitigation strategies}

\author{L. D. Valdez} 

\affiliation{Departamento de F\'{i}sica, Facultad de Ciencias Exactas y
  Naturales, Universidad Nacional de Mar del Plata, and Instituto de
  Investigaciones F\'{\i}sicas de Mar del Plata (IFIMAR-CONICET), De\'an
  Funes 3350, 7600 Mar del Plata, Argentina}

\author{H. H. Arag\~ao R\^ego}

\affiliation{Departamento de F\'{\i}sica, Instituto Federal de
  Educa\c{c}\~ao, Ci\^{e}ncia e Tecnologia do Maranh\~ao, S\~ao
  Lu\'{\i}s, MA, 65030-005, Brazil}

\author{H. E. Stanley} 

\affiliation{Center for Polymer Studies, Boston University, Boston,
  Massachusetts 02215, USA.}

\author{L. A. Braunstein} 

\affiliation{Departamento de F\'{i}sica, Facultad de Ciencias Exactas y
  Naturales, Universidad Nacional de Mar del Plata, and Instituto de
  Investigaciones F\'{\i}sicas de Mar del Plata (IFIMAR-CONICET), De\'an
  Funes 3350, 7600 Mar del Plata, Argentina}

\affiliation{Center for Polymer Studies, Boston University, Boston,
  Massachusetts 02215, USA.}

\begin{abstract}

The Ebola virus is spreading throughout West Africa and is causing
thousands of deaths. In order to quantify the effectiveness of different
strategies for controlling the spread, we develop a mathematical model
in which the propagation of the Ebola virus through Liberia is caused by
travel between counties. For the initial months in which the Ebola virus
spreads, we find that the arrival times of the disease into the counties
predicted by our model are compatible with World Health Organization
data, but we also find that reducing mobility is insufficient to contain
the epidemic because it delays the arrival of Ebola virus in each county
by only a few weeks. We study the effect of a strategy in which safe
burials are increased and effective hospitalisation instituted under two
scenarios: (i) one implemented in mid-July 2014 and (ii) one in
mid-August---which was the actual time that strong interventions began
in Liberia.  We find that if scenario (i) had been pursued the lifetime
of the epidemic would have been three months shorter and the total
number of infected individuals 80\% less than in scenario (ii). Our
projection under scenario (ii) is that the spreading will stop by
mid-spring 2015.

\end{abstract}


\maketitle

\section*{INTRODUCTION}

{\it For a fleeting moment last spring, the epidemic sweeping West
  Africa might have been stopped. But the opportunity to control the
  virus, which has now caused more than 7,800 deaths, was lost\/} \cite{NYT}.

The current Ebola outbreak in Western Africa is one of the deadliest and
most persistent of epidemics \cite{Fri_01}. According to World Health
Organization data \cite{WHO_01} as of 31 December 2014 there have been
20,171 cases and 7,889 deaths in three countries alone: Guinea, Sierra
Leone, and Liberia. These numbers increase when cases and deaths from
countries in which the outbreak has been officially declared over
\cite{CDC_01} are included.

Cultural, economic, and political factors in that region of Western
Africa \cite{Fri_01, ALJ_01, GUA_01, GLO_01, WHA_01, WOR_01} have
hampered the effectiveness of the intervention strategies used by the
health authorities.  Because of a lack of reliable information about
local patterns of the spreading of the Ebola virus disease
(EVD) \cite{SCI_01, N24_01, NZE_01}, the strategies currently being
used, including the mobilisation of resources, the creation of new Ebola
treatment centers (ETC), the development of safe burial procedures, and
the international coordination of the efforts \cite{WHO_02} as of 1
January 2015 have been only partially successful.

Legrand et al.~\cite{Leg_01} developed a seminal mathematical
stochastic model with full mixing that reproduces the 1995 EVD
outbreak in the Congo and the 2000 outbreak in Uganda. The population
is divided into six compartments.  Individuals in the susceptible
compartment transition to exposed compartment and to the infectious
compartment when they become infected. A percentage of these infected
individuals are hospitalised and there are two possible outcomes: (i)
they die, but before they are removed from the epidemic system they
transition into the funeral compartment and infect other susceptible
individuals, or (ii) they are removed from the system because they are
cured.  The maximum likelihood method is used to calibrate the model
with the data.

Rivers et al.~\cite{Riv_01} used a deterministic version of this model
and least-squares optimisation to fit the current Liberia and Sierra
Leone outbreak data. Their model indicated that the epidemic would not
reach its peak until 31 December 2015.  Gomes et al.~\cite{Gom_01}
estimated the transmission coefficients using the model provided by
Ref.~\cite{Leg_01}, a Global Epidemic and Mobility model that uses a
structured metapopulation scheme, integrating the stochastic modelling
of the disease dynamic, high resolution census and human mobility
patterns at the global scale using a high resolution population
data~\cite{Balcan_01,Balcan_02}. The parameters were estimated by
fitting the total number of cumulative deaths from Liberia, Sierra
Leone, and Guinea during the period 6 July -- 9 August 2014. The
transmission parameters obtained were used to forecast three months of
EVD propagation in West Africa and the probability of its spreading
internationally. They found that the risk of cases spreading to other
countries was low.  Poletto $et\;al.$ \cite{Pol_01} used the same
model and found that reducing the number of travellers crossing
international boundaries delays the arrival of EVD by only a few
weeks. Merler $et\;al.$ \cite{Mer_01} used methods similar to those in
Ref.~\cite{Gom_01} to model the effect of epidemic spreading between
geographical regions. They took into account the movements of
non-infected individuals who were assisting in health-care facilities,
those who took care of non-hospitalised infected individuals, and
those who attended funerals.

Population mobility---the movement of individuals seeking safer areas,
better health infrastructures, or food supplies---strongly affects
disease propagation and plays a major role in epidemic spreading and
in the effectiveness of any intervention scheme \cite{Wes_01}.  In
Liberia, 54\% of the population over the age of 14 are internally
displaced \cite{LIBS_01}. Understanding these patterns of movement is
essential when planning interventions to contain regional outbreaks.
In recent years a number of mobility studies have been published
\cite{Gon_01, Tat_01, Wes_02}, including Wesolowski et
al.~\cite{Wes_01}, who used mobile telephone network data to analyse
mobility patterns that could be useful to understand the Ebola
outbreak. They analysed data sources from mobile phone call records
(CDRs), national census microdata samples, and spatial population data
in order to estimate domestic and international mobility patterns in
West African countries. The best mobility estimates were obtained for
Senegal, Cote d'Ivoire, and Kenya, and Wesolowski {\it et al.}
\cite{Wes_01,Wes_02} used them to produce a spatial interaction model
of national mobility patterns in order to estimate how the EVD
affected regions are connected by population flows.

We use a stochastic compartmental model and a set of differential
equations, which are the quasi-deterministic representation of a
stochastic model, to understand how population mobility affects the
spreading of EVD between regions (counties) within Liberia.  Our model
quantifies how mobility between counties affects epidemic spreading
inside Liberia, and we find that although reducing mobility among
counties delays the spread of Ebola, it does not contain it. Our study
indicates that the response implemented in August 2014 will result the
extinction of the epidemic by mid-spring 2015, but it also indicates
that an earlier response would have been extremely effective in
containing the disease.

\section*{RESULTS}

\subsection*{Model}

In our model we classify individuals as susceptible (S), exposed (E),
i.e., infected but not infectious, infected (I), hospitalised (H),
recovered (R), i.e., either cured or dead with a safe burial that does
not transmit the disease, or dead (F) with an unsafe burial that
transmits the disease. We also classify infected and hospitalised
individuals according to their fate: those who are infected, will be
hospitalised, and will die (${\rm I_{\rm DH}}$), those who are infected,
won't be hospitalised, and will die (${\rm I_{\rm DNH}}$), those who are
infected, will be hospitalised, and will recover (${\rm I_{\rm RH}}$),
those who are infected, won't be hospitalised, and will recover (${\rm
  I_{\rm RNH}}$), those who are hospitalised and will die (${\rm H_{\rm
    D}}$), and those who are hospitalised and will recover (${\rm H_{\rm
    R}}$). The symbols S, E, I, H, R, and F indicate both the
classification and the population percentage within the classification.

\begin{figure}
\includegraphics[scale=0.75]{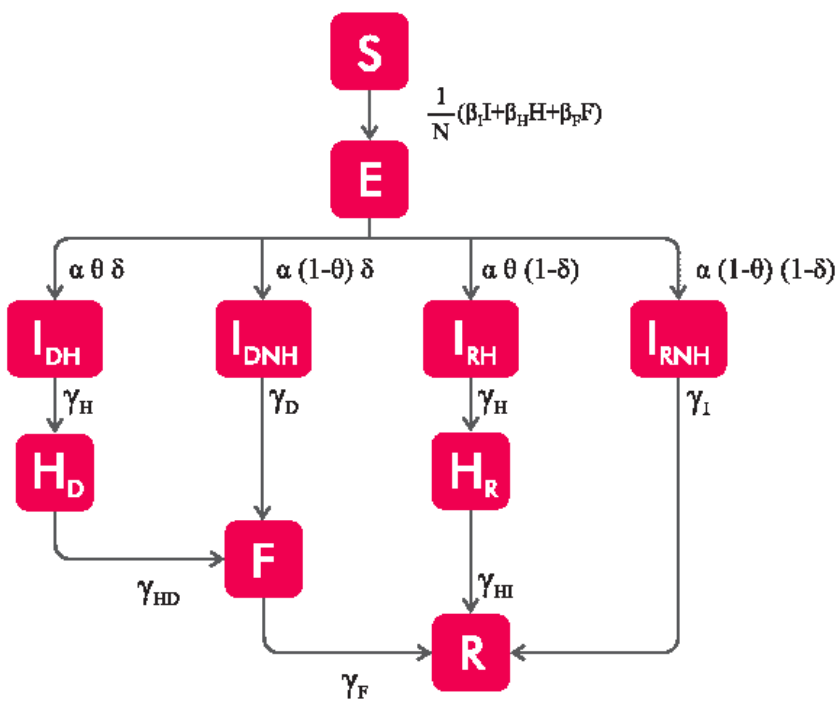}
\bigskip
\caption{{ \bf A schematic of the transitions between different states
    of our model for the EVD spreading in West Africa 2014 and their
    respective transition rates.}  In the model, the population is
  divided into ten compartmental states (See Table~S1): Susceptible
  ($S$) individuals who in contact with infected individuals can become
  exposed ($E$). These $E$ individuals after the incubation period
  become infected and follow four different scenarios: (i) Infected
  individuals that will be cured ---recovered--- without hospitalisation
  (${\rm I_{\rm RNH}}$); (ii) Infected individuals who will be cured
  (${\rm I_{\rm RH}}$) after spending a period on a hospital ($H_R$);
  (iii) Infected individuals without being hospitalised (${\rm I_{\rm
      DNH}}$) who will die and may infect other individuals in their
  funerals ($F$); and (iv) Infected individuals (${\rm I_{\rm DH}}$)
  that even after spending a period in a hospital ($H_D$) will die and
  may also spread the infection in the funerals ($F$). Recovered
  individual ($R$), can be cured or dead.
\label{f.rules}}
\end{figure}

Figure~\ref{f.rules} shows a schematic presentation of the model
indicating the compartmental states (red boxes) and the transition rates
among the states (connecting arrows). The $I$,
$I=I_{DH}+I_{DNH}+I_{RH}+I_{RNH}$ represents the total number of
infected individuals, and $H=H_{R}+H_{D}$ the total number of those
hospitalised. Table~\ref{table.2} shows the different parameters used to
calculate the transition rates among the different compartmental states,
and Table~S1 (see Supplementary Information) shows the $N_{T}=12$
transitions between states and their rates $\lambda_i$ with $i=1,
...N_T$.

To determine how geographic mobility spreads the disease, we utilise
the model of West African regional transportation patterns developed
by Wesolowski $et\; al.$ \cite{Wes_01,Wes_02}. In their research they
applied a gravity model to mobile phone data for Senegal to estimate
the flow of individuals between counties in Liberia. Although these
movement data are ``historical'' and do not reflect how local
population behaviour may have changed in response to the current
crisis, we assume the patterns of mobility obtained in the Wesolowski
model \cite{Wes_01,Wes_02} still represent a good approximation of the
routine commuting patterns of the population in Liberia prior to the
outbreak. This is different to post-outbreak models that describe
travel patterns that reflect human efforts to avoid the disease or to
attend funerals of epidemic victims (see Ref.~\cite{Mer_01} and
references in therein).

We assume that there is a flow of individuals between all $N_{\rm
  co}=15$ counties of Liberia, and that only susceptible or exposed
individuals can travel between counties. Thus the deterministic
evolution equations for the number of individuals in each state in
county $c$ in our model are
\begin{eqnarray}\label{Eq.evol}
 \frac{d\;S^c}{d\;t}&=& -\frac{1}{N_c} (\beta_I S^c I^c + \beta_H S^c
 H^c +\beta_F S^c F^c)+ \sigma_c (\bar{S})\; ;\label{ecu.1}\\ 
 \frac{d\;E^c}{d\;t}&=& \frac{1}{N_c} (\beta_I S^c I^c + \beta_H S^c H^c
 + \beta_F S^c F^c) - \alpha E^c+  \sigma_c (\bar{E})\; ;\\ 
 \frac{d\;I^c_{DH}}{d\;t}&=& \alpha \;\delta \;\theta E^c  - \gamma_H
 \;I^c_{DH} ;\\ 
 \frac{d\;I^c_{DNH}}{d\;t}&=& \alpha \;\delta \;(1-\theta)\; E^c  -
 \gamma_D \;I^c_{DNH};\\ 
 \frac{d\;I^c_{RH}}{d\;t}&=& \alpha \;(1-\delta) \;\theta E^c  -
 \gamma_H \;I^c_{RH} ;\\ 
 \frac{d\;I^c_{RNH}}{d\;t}&=& \alpha \;(1-\delta) \;(1- \theta)\; E^c  -
 \gamma_I \; I^c_{RNH}  ;\\ 
 \frac{d\;H^c_{D}}{d\;t}&=& \gamma_H \;I^c_{DH}  - \gamma_{HD} \;H^c_{D} \; ; \\
 \frac{d\;H^c_{R}}{d\;t}&=& \gamma_H \;I^c_{RH}  - \gamma_{HI} \;H^c_{R} \; ;\\
 \frac{d\;F^c}{d\;t}&=& \gamma_D \;I^c_{DNH}  + \gamma_{HD} \;H^c_{D}- \gamma_F \;F^c;\\
 \frac{d\;R^c}{d\;t}&=& \gamma_I \;I^c_{RNH}  +\gamma_{HI} \;H^c_{R} +
 \gamma_F \;F^c, \label{ecu.10} 
\end{eqnarray}
where $\sigma_c$ is the total rate of mobility in each county $c$ and
is given by
\begin{equation}\label{mobility}
\sigma_c(\bar{x})= \sum_{c \neq j} \frac{x^j}{N_j} r_{j\; c}
- \frac{x^c}{N_c} \sum_j r_{c \;j} ,
\end{equation}
where $x^j$ ($x^c$) is the number of individuals (susceptible or
exposed), in county $j$ ($c$), $N_j$ ($N_c$) the total population of
county $j$ ($c$), and $r_{j\; c}$ and $r_{c\; j}$ the mobility rates
from county $j \to c$ and from county $c \to j$, respectively. Note
that due to mobility the population in each county changes, but since
this evolution is much slower than the dynamics of the disease
spreading, we consider $N_c$ to be constant (in our model without
restriction on the mobility, the population in each county changes
less than a 5\% each year). In addition, in this model we disregard
the mobility inside each county, i.e., we assume that the population
is fully mixed. Because recovered individuals are unable to transmit
the disease or be reinfected, they do not affect the results of our
model and we disregard their movements between counties.

In the context of complex network research, this model of mobility
between counties breaks the traditional full-mixing approach because
each county can be thought of as a node of a metapopulation network
\cite{Col_01} in which the weight of each link is proportional to the
mobility flow. Note that if in Eqs.~(\ref{ecu.1})--(\ref{ecu.10}) we
drop the index $c$ and disregard the flow mobility we are no longer
taking the counties into account, and we have a scenario that represents
the spread throughout the entire country.

\subsection*{Transmission rates estimated}

\begin{figure}[H]
\centering
\includegraphics[scale=0.25]{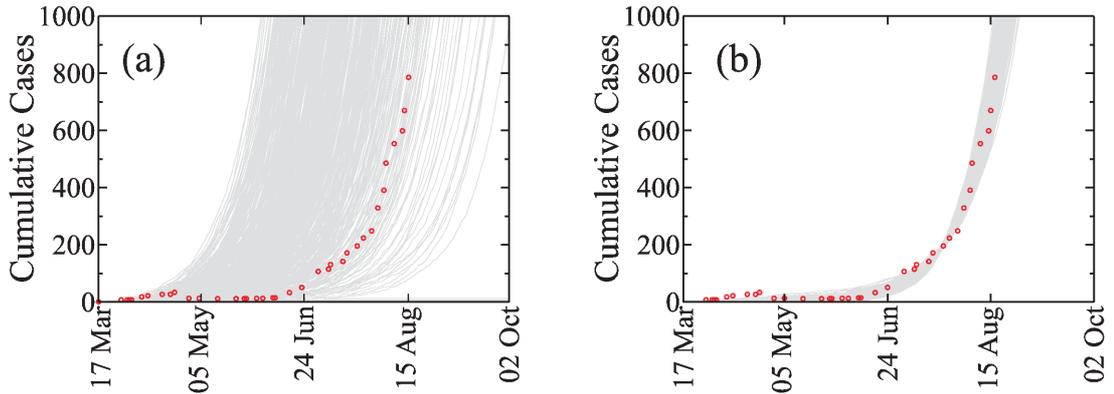}
\caption{{\bf Cumulative number of cases in Liberia for the parameters
    given in Table~\ref{table.2}}. Cumulative number of cases obtained
  with our stochastic model with the transition presented in Table~S1
  and Eq.~(\ref{mobility}) in Liberia with $1000$ realisations (gray
  lines) and the data (symbols) without temporal shift (a) and (b) with
  a temporal shift using $s_c=200$.  The transmission coefficients
  $\beta_I=0.14$, $\beta_H=0.29$ and $\beta_F=0.40$ were obtained as
  explained in {\it Methods: Calibration with the deterministic
    equations}. From the WHO's data the index case is located at Lofa on
  March 17 2014.
\label{f.Liberia}}
\end{figure}

According to WHO data \cite{WHO_01}, the first index case (patient zero)
was diagnosed in Lofa on 17 March 2014. Thus our initial conditions in
Lofa are (i) one infected individual in that county and (ii) the rest of
the population susceptible.  The estimated rates of transmission in
day$^{-1}$ obtained (using the method presented in the section {\it
  Methods: Calibration with the deterministic equations}~\ref{s.method})
are $\beta_I=0.14\;[0,0.26]$ in the community, $\beta_H=0.29\;[0,0.92]$
in the hospitals, and $\beta_F=0.40\;[0,0.99]$ at the funerals, where
the intervals correspond to the values used to obtain the average rates
of transmission obtained from the Akaike criterion.  From these rates,
we construct the next-generation matrix \cite{Van_01,Die_01} (see {\it
  Methods: Estimation of $R_0$}) in order to compute the reproductive
number $R_0$, defined as the average number of people in a susceptible
population one infected individual infects during his or her infectious
period. This parameter is fundamental when predicting whether a disease
can reach a macroscopic fraction of individuals \cite{And_01}. For a
critical value $R_0=1$ there is a phase transition below which no
epidemic takes place, and the disease is only a small outbreak, while
for $R_0>1$ the probability that an epidemic spreading develops is
greater than zero \cite{And_01}. For the values of rates of transmission
given above, we find that the reproductive number of the current EVD
outbreak is $R_0=2.11\; [1.88,2.71]$, well above the critical threshold
$R_0= 1$, where the interval was obtained from the transmission
coefficient selected from the Akaike criterion. This value of $R_0$ is
compatible with the one obtained in Ref.~\cite{Riv_01}. We run our
stochastic simulations presented in {\it Methods: Stochastic model\/}
for these estimated values in order to compare the total number of cases
with the data given by WHO before the interventions began in the middle
of August 2014 \cite{WHO_02}. Figure~\ref{f.Liberia}(a) plots the number
of cumulative cases as a function of time for 1000 realisations of our
stochastic model and compares the results with WHO data \cite{WHO_01}
without any shift correction. The individual realisations have the same
shape as the data but due to the stochasticity at the beginning of the
outbreak the exponential increase in the number of cases occurs at
different moments.

Figure~\ref{f.Liberia}(b) plots the cumulative number of cases as a
function of time with the initial conditions explained above when a
temporal shift is applied to the stochastic simulations. The agreement
between the simulations and the data indicates that our model can
successfully represent the dynamics of the spreading of the current
Ebola outbreak in Liberia.

\subsection*{The geographical spread of Ebola cases across Liberia due
  to mobility} 

\begin{figure}[H]
\centering
\includegraphics[scale=0.25]{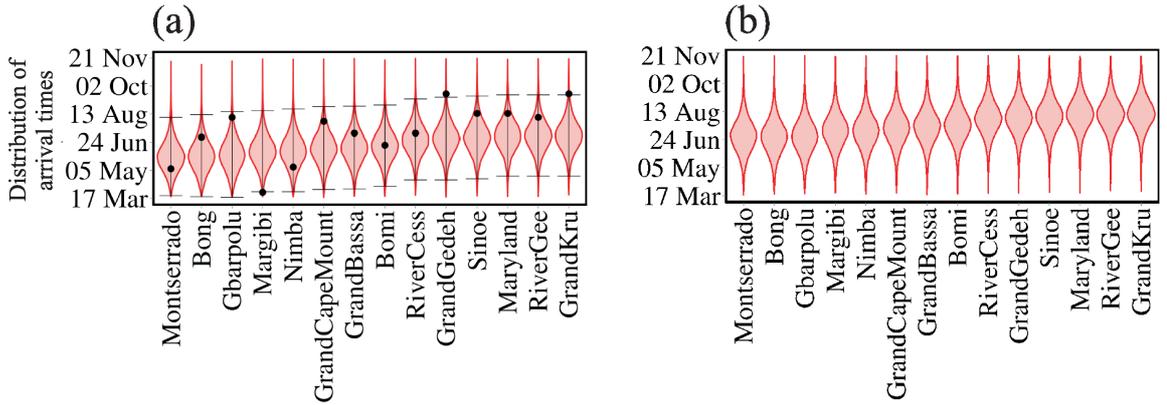}
\caption {Figure a: Violin plots representing the distribution of
  arrival time $t_a$ to each county considering the mobility flow of
  individuals among counties \cite{Wes_01} without any restriction on
  the mobility. The results are obtained from our stochastic model with
  the estimated transmission coefficients over $1000$
  realisations. Error bars indicate the 95\% confidence interval. From
  WHO's reports the index case (patient zero) was located at Lofa at 17
  of March 2014. The circles represent the values of $t_a$ reported by
  WHO. The very early case of Margibi is below the 5\% probability, and
  it is explained in Ref.~\cite{Exxon_02}.  Figure b: Violin plots
  representing the distribution of arrival time $t_a$ to each county
  reducing the mobility among counties by 80\%.}
\label{f.migration_Liberia}
\end{figure}

The mobility among the 15 counties allows us to compute the arrival time
$t_a$ in each county, assuming that the index case was in Lofa on 17
March 2014. Figure~\ref{f.migration_Liberia} shows the violin plots of
the arrival times $t_a$ of the disease as it spreads from Lofa County
into the other 14 Liberian counties and compares our results with those
supplied in the WHO reports (circles).

Comparing the results of our predictions of the arrival times of the
first case as it spreads to the other counties with the WHO data (see
Fig.~\ref{f.migration_Liberia}a), all counties except Margibi and Grand
Gedeh fall into a 95\% confidence interval. This could be caused by (i)
an underestimation of the number of cases in the WHO data \cite{WHO_01}
due to a lack of information \cite{NYT}, or (ii) because the data
recorded are actually the times of reporting and not the times of onset.

As the disease began to spread, population mobility decreased. This was
in part due to imposed regulations attempting to contain the disease but
also due to the population's fear of contagion. We reflect this in our
model by decreasing the mobility value.
Figure~\ref{f.migration_Liberia}(b) shows the arrival times produced by
our model when, as a strategy for slowing the spread, the mobility is
reduced by 80\%. Note that this reduction delays the arrival of EVD in
each county by only a few weeks. This suggests that reducing the
mobility of the individuals between counties will not stop the spread
but will slow it sufficiently that other strategies can be developed and
applied.  Reducing mobility is also insufficient when considering
international transmission of the disease \cite{Pol_01} and more
aggressive interventions are needed. We believe that an increase in both
the percentage of infected individuals receiving hospitalisation in ETCs
and the percentage of burials that follow procedures that do not
transmit the disease are essential in containing the epidemic.

\subsection*{Interventions and time to extinction} 

To contain the disease and reduce its transmission we reduce mobility
by 80\%, increase the number of burials following procedures that do
not transmit the disease, and increase the rate of hospitalisation in
ETCs.  Because health workers in ETCs have specialised training, we
assume that the probability that they will be infected is greatly
reduced and that the transmission coefficient $\beta_H$ is decreased.
A sufficiently rapid response to the EVD by the ETCs requires that
$\beta_H$ be decreased exponentially to a final value of $10^{-3}$,
and hospitalisation $\theta$ must be increased exponentially to reach
$\theta = 1$. On the other hand, when changing local burial customs we
assume that $\beta_F$ decreases linearly and approaches zero. Changing
local burial customs involves recruiting and training burial teams,
takes a longer period of time, and is less aggressive than other kinds
of intervention.  This approach allows us to estimate an upper limit
for the end of the epidemic, because we did not take into account
other measures applied, such as, contact tracing, which could make the
estimated end to the epidemic occur earlier.

We apply these changes to simulate a two-month period, and the final
result is that $R_0$ decreases from 2.11 to $R_{0}=0.69$, which is below
the epidemic threshold.  We consider two scenarios, (i) implementing the
strategy beginning August 15 (the middle of the month indicated by WHO
\cite{Exxon_03} for the outbreak of EVD in West Africa) in which all
symptomatic individuals are admitted to ETCs and safe burial procedures
begin to apply, or (ii) implementing the same strategy, but beginning
July 15 in order to study how delaying the implementation of the
strategies affected containment. Our goal is to demonstrate that if the
international response had been more rapid, the spreading disease would
have been contained with a 50\% probability by early March 2015 instead
of the end of May 2015.

\begin{figure}[H]
\centering
\includegraphics[scale=1]{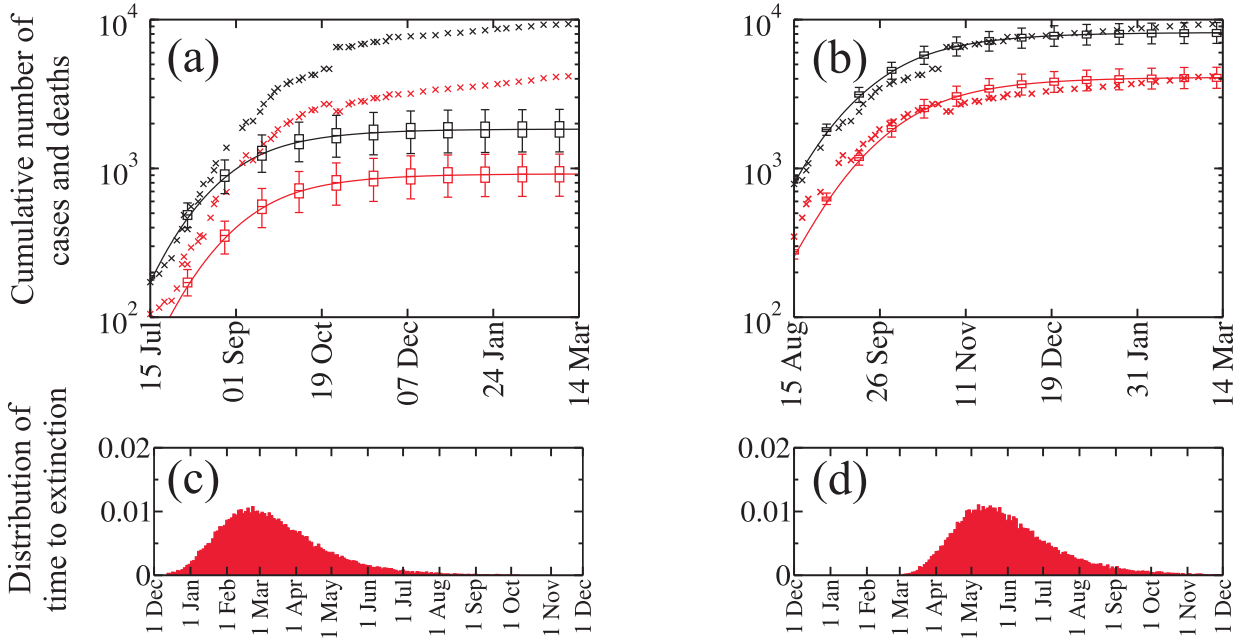}
\caption{Evolution of the number of cases (black) and deaths (red)
  when a reduction of 80\% in the mobility rates is applied.  The
  value of $\beta_H$ decreases exponentially to reach the value
  $10^{-3}$ and $\beta_F$ decreases linearly to reach a 0\% of their
  original values. Also the hospitalisation fraction increases
  exponentially to reach $\theta = 1$. All reductions in the
  transmission coefficient were applied during two months, for (a)
  beginning at July 15$th$ and (b) August 15$th$. Solid lines were
  obtained from the evolution equations~(\ref{ecu.1}-\ref{ecu.10}) and
  the symbols are the data. The box plots show the median, the 25th and
  75th percentile and 95\% confidence interval of the median, obtained
  from the stochastic simulations. Figures c) and d) are the
  distribution of time to extinction of the EVD epidemic obtained from
  the stochastic simulations, when the strategy is applied from middle
  July and from middle August 2014, respectively. We show these
  distributions from December 1st, 2014 to December 1st, 2015.}
\label{strategyFunHospMigrat}
\end{figure}

Figures~\ref{strategyFunHospMigrat}(a) and
\ref{strategyFunHospMigrat}(b) show that reducing the number of cases
produced in hospitals and funerals reduces the cumulative number of
cases to a plateau lower than the one predicted when no strategies are
applied \cite{Mel_01}.  Figure~\ref{strategyFunHospMigrat}(a) shows
that if our strategy had been applied in the middle of July the
cumulative number of cases and deaths would have been approximately
80\% lower than the reported number that resulted when the strategies
were instituted in the middle of
August. Figure~\ref{strategyFunHospMigrat}(b) shows that when we apply
the strategy of our model to the actual mid-August starting time, it
predicts (between the 95\% confidence interval of the median) the actual
trend of cases and deaths reported in the WHO data in mid-March 2015.

Our stochastic model allows us to quantify how the two different
strategy implementation times affect the extinction time of the EVD
epidemic.  Figures~\ref{strategyFunHospMigrat}(c) and
\ref{strategyFunHospMigrat}(d) show the extinction time distributions,
i.e., when $E=I=H=F=0$, when the strategy is implemented in July 2014
and August 2014, respectively (for the initial conditions, we use the
cases provided by WHO for these dates). We find that the median of this
distribution when the strategy is implemented in July is 6 March 2015
(with a 95\% confidence interval from 5 January 2015 to 1 July 2015) and
when it is implemented in August is 25 May 2015 (with a 95\% confidence
interval from 28 March 2015 to 20 September 2015).  Implementation in
mid-August generated 8,000 cases of the disease, but an implementation
in mid-July would have reduced the time to disease extinction by three
months and generated only 1,700 cases. The mid-August implementation
faced a larger number of cases, the disease progression had a greater
inertia against the strategy, and the cumulative number of cases
required a longer time to go from an exponential regime to a
subexponential regime.  Thus if the health authorities and the
international community had acted sooner the number of infected people
would have been much lower.

\section*{DISCUSSION}

In this manuscript we study the spreading of the Ebola virus using
stochastic and deterministic compartmental models that incorporate the
mobility of individuals between the counties in Liberia.  We find that
our model describes well the arrival of the disease into each of the
counties, that reducing population mobility has little effect on
geographical containment of the disease, and that reducing population
mobility must be accompanied by other intervention strategies. We thus
examine the effect of an intervention strategy that focuses on both an
increase in safer hospitalisation and an increase in safer burial
practices. Our study indicates that the intervention implemented in
August 2014 reduced the total number of infected individuals
significantly when compared to a scenario in which there is no strategy
implementation, and it predicts that the epidemic will be extinct by
mid-spring 2015. We also use our model to consider the difference in
outcome had the strategy been implemented one month earlier. We find
that the cumulative number of cases and deaths would have been
significantly lower and that the epidemic would have ended three months
earlier. This indicates that a rapid and early intervention that
increases the hospitalisation and reduces the disease transmission in
hospitals and at funerals is the most important response to any possible
re-emerging Ebola epidemic.

Although our model simplifies the dynamics of epidemic spreading, it
provides an adequate picture of the evolution in the number of cases
and deaths. In future research we will incorporate more aspects of
population mobility and intervention strategies carried out by health
authorities. This will enable us to describe in greater detail the
evolution of an epidemic and the efficacy of different strategies. 

Finally, the methods used in this manuscript to study Liberia can also
be applied to Guinea and Sierra Leone as soon as high quality epidemic
data from those countries become available. Future work should include
both countries in order to quantify the cases spreading from them into
Liberia.

\section*{METHODS}\label{s.method}
\subsection*{Stochastic model}\label{s.methodSto}

We generate a stochastic compartmental model based on the Gillespie
algorithm.  At each iteration of the simulation we draw a random number
$\tau$ (which represents the waiting time until the next transition)
from an exponential distribution with parameter $\Delta$ given by
parameter
\begin{equation}
\Delta=\sum_{i=1}^{N_T}\sum_{j=1}^{N_{\rm co}}
\lambda_{i}^{j}+\sum_{i=1}^{N_{\rm co}}\sum_{j=1}^{N_{\rm co}} (E^j+S^j)r_{j\;i}/N_j.
\end{equation}
Here the first term $\lambda_i^{j}$ is the rate of transition between
states $i$ in county $j$ given in Table~S1, and the second term
corresponds to the mobility rates given in Eq.~(\ref{mobility}) with
$x=E$ and $x=S$.

\subsection*{Calibration with the deterministic equations}\label{s.methodCal}

To estimate the transmission coefficients $\beta_I$, $\beta_H$, and
$\beta_F$ we calibrate a system of differential equations using
least-squares optimisation with the data of the total cases from
Liberia in the March--August period \cite{WHO_01}, and we apply a
temporal shift, which we will explain below. We compute the
least-square values using a set of parameters generated using Latin
hypercube sampling (LHS) in the parameter space $[0,1]^3$, which we
divide into $10^6$ cubes of the same size. For each cube we choose a
random point as a candidate for $(\beta_I,\beta_H,\beta_F)$ in order
to compute the standard deviation between the data and the system of
differential equations obtained from this point.  At the beginning of
the epidemic there are very few cases (infected individuals), thus the
evolution of the disease is in a stochastic regime in which the
dispersion of the number of new cases is comparable to its mean value
(see Refs.~\cite{Vol_01,Val_01,Bar_01}). When the number of infected
individuals increases to a certain level, however, the epidemic
evolves toward a quasi-deterministic regime and the evolution of the
states of the stochastic simulation is the same as the states obtained
using the solution of the evolution equations
(Eqs.~\ref{ecu.1}-\ref{ecu.10}). Nevertheless, due to fluctuations in
the initial stochastic regime, a random temporal displacement of the
quasi-deterministic growth of the number of accumulated cases is
generated.  Thus to remove this stochastic temporal shift and to
compare the three aspects---the simulations, the numerical solution,
and the data---we set the initial time at $t = 0$ when the total
number of cases is above a cutoff $s_c$ \cite{Vol_01,Val_01,Bar_01}.
For the calibration of the transmission coefficients, we use $s_c=200$
(which corresponds to the cumulative number of cases after 21 July,
according to the WHO data \cite{WHO_01}), and using the least square
method we give the data above this cutoff 50\% of the weight because
we are assuming that above $s_c$ the evolution of the disease
spreading is quasi-deterministic. Finally, after we compute the sum of
square residuals for each point in the parameter space, we apply the
Akaike information criterion (AIC) and average those candidates of
$(\beta_I,\beta_H,\beta_F)$ with a AIC difference $\Delta<2$
\cite{Bur_01} to obtain a model-averaged estimate of the transmission
coefficients. An alternative method for estimating the transmission
coefficients using an exponential fitting is discussed in {\it
  Supplementary Information: Calibration}. We find that this fitting
generates the same set of values of transmission coefficients than the
method with a temporal shift $s_c$.  Additionally, in {\it
  Supplementary Information: Sensitivity Analysis} we discuss the
sensitivity of the estimated values of the transmission coefficients
when $\theta$ and $\delta$ change.

The mobility data for the Wesolowski model were provided by Flowminder
\cite{Wes_01,FLOW_01,WPOP_01} and the {\it total cumulative\/} case data
used to calibrate the model were those supplied in reports generated by
WHO \cite{WHO_01}. Note that in this work we do not calibrate the model
to cases in each county because it was shown by Chowell et
al. \cite{Cho_01} that globally the number of cases grows exponentially
but locally can be better approximated by a polynomial than by an
exponential growth. This cannot be addressed using our model because
mathematically a differential equation with constant rates only
reproduces an exponential growth.

Weitz and Dushoff \cite{Wei_01} recently demonstrated that calibration
causes an identification problem, i.e., that many combinations of the
coefficient transmission values reproduce the real evolution of the
number of cases, which is compatible with our finding that the
calibrated transmission coefficients are in a plane (see Supplementary
Information: Calibration). This point should be addressed in future
research.

\subsection*{Estimation of $R_o$}\label{s.methodR0}

In order to compute the reproduction number $R_{0}$, following van den
Driessche et al.~\cite{Van_01} and Diekmann et al.~\cite{Die_01}, we
construct a next-generation matrix.

First, using the Jacobian matrix of the system of Eqs.~(1-10) we
construct the ``transmission matrix'' {\bf F}, and the ``transition
matrix'' {\bf V} , obtaining
$$\bf{F}=\bordermatrix{\text{}&&&&&\cr
                &\bf{F_1} &  \bf{0} &  \bf{0} & \ldots & \bf{0}\cr
                & \bf{0}  &  \bf{F_2}&  \bf{0} & \ldots & \bf{0}\cr
                & \bf{0}  &  \bf{0}&  \bf{F_3} & \ldots & \bf{0}\cr
                & \vdots & \vdots & \vdots& \ddots & \vdots\cr
                & \bf{0}  &   \bf{0} &  \bf{0}      &\ldots & \bf{F_{15}}}$$
where
$$\bf{F_i}=\bordermatrix{\text{}&&&&&&&&\cr
                &0 &  \beta_I & \beta_I &\beta_I&\beta_I&\beta_H&\beta_H&\beta_F\cr
                &0  & 0 & 0 & 0&0  & 0 & 0 & 0\cr
                & \vdots & \vdots& \vdots & \vdots& \vdots & \vdots& \vdots & \vdots\cr
               &0  & 0 & 0 & 0&0  & 0 & 0 & 0}$$
and
$$\bf{V}=\bordermatrix{\text{}&&&&&\cr
                &\bf{V_{1,1}} &  \bf{V_{1,2}} &  \bf{V_{1,3}} & \ldots & \bf{V_{1,15}}\cr
                & \bf{V_{2,1}}  & \bf{V_{2,2}} &  \bf{V_{2,3}}  & \ldots & \bf{V_{2,15}} \cr
                & \vdots & \vdots & \vdots& \ddots & \vdots\cr
                & \bf{V_{15,1}}   & \bf{V_{15,2}} & \bf{V_{15,3}} &\ldots & \bf{V_{15,15}}}$$
where
$$\bf{V_{i,i}}=\bordermatrix{\text{}&&&&&&&&\cr
                &\alpha+\sum_u r_{iu}/N_{i} & 0 &0 &0&0&0&0&0\cr
                &-\alpha\delta\theta  & \gamma_H & 0 & 0&0  & 0 & 0 & 0\cr
                &-\alpha\delta(1-\theta)  & 0& \gamma_D & 0&0  & 0 & 0 & 0\cr
                &-\alpha(1-\delta)\theta  & 0& 0 & \gamma_H&0  & 0 & 0 & 0\cr
                &-\alpha(1-\delta)(1-\theta)  & 0& 0 & 0&\gamma_I  & 0 & 0 & 0\cr
                &0  & -\gamma_H& 0 & 0&0  & \gamma_{HD} & 0 & 0\cr
                &0  & 0& 0 & -\gamma_{H}&0  & 0 & \gamma_{HI} & 0\cr
                &0  & 0& -\gamma_D & 0&0  & -\gamma_{HD} & 0 & \gamma_{F}}$$
$$\bf{V_{i,j}}=\bordermatrix{\text{}&&&&&&&&\cr
                &-r_{ji}/N_{j} &  0 & 0 &0&0&0&0&0\cr
                &0  & 0 & 0 & 0&0  & 0 & 0 & 0\cr
                & \vdots & \vdots& \vdots & \vdots& \vdots & \vdots& \vdots & \vdots\cr
               &0  & 0 & 0 & 0&0  & 0 & 0 & 0}$$ with $i\neq j$.

Note that the mobility rates are only in the transition matrix.  Using
these matrices we construct the next generation matrix, defined as ${\bf
  FV}^{-1}$. Finally, the reproduction number is given by the spectral
ratio $\rho$ of the next generation matrix, $R_{0}=\rho({\bf FV}^{-1})$,
i.e., its highest eigenvalue. Note that when the mobility rates go to
zero, $R_0$ decreases, i.e., in this limit an infected individual in a
given county cannot interact with people from other counties and can
only transmit the disease to susceptible individuals in the same county.


\bigskip

\begin{table}[H]
\begin{tabular}{|l|l|l|}
  \hline
  Transition Parameters  & Value & References \\\hline
  Mean duration of the incubation period $(1/\alpha)$ & $7$ days &  \cite{Bwa_01,Nda_01,Dow_01}\\
  Mean time from the onset to the hospitalisation ($1/\gamma_H$) & $5$ days & \cite{Kha_01}  \\
  Mean duration from onset to death ($1/\gamma_D$)& $9.6$ days&  \cite{Kha_01} \\
  Mean time from onset to the end for the cured ($1/\gamma_I$)& $10$ days & \cite{Dow_01,Row_01} \\
  Mean time from death to traditional burial ($1/\gamma_F$)& $2$ days & \cite{Leg_01} \\
  Proportion of cases hospitalised ($\theta$)& 50\%& \cite{Riv_01} \\
  Fatality Ratio ($\delta$)& 50 \% & \cite{Riv_01} \\
  Mean time from hospitalization to end for cured ($1/\gamma_I$)& $5$ days  &  \cite{Leg_01} \\
  Mean time from hospitalization to dead ($1/\gamma_{HD}$)& $4.6$ days &  \cite{Leg_01}\\\hline
\end{tabular}
\caption{{\bf Transition parameters used to calculate the transition
    rates in our epidemic model.} Table describing the different
  parameters used to calculate the transition rates among the ten
  different compartmental states in our model.}\label{table.2}
\end{table}

\bigskip
\newpage

\renewcommand{\thefigure}{S\arabic{figure}}
\renewcommand{\theequation}{S\arabic{equation}}
\renewcommand{\thetable}{S\arabic{table}}
\renewcommand{\thesection}{\Roman{section}}
\setcounter{table}{0}
\setcounter{figure}{0}
\setcounter{equation}{0}

\section*{SUPPLEMENTARY INFORMATION}

\subsection*{Calibration}

In the main text, using the least square method and a temporal shift
for the number of cases ($s_c=200$) we obtained the coefficient
parameters $\beta_I$, $\beta_H$ and $\beta_F$. We can also obtain
these transmission coefficients by measuring the parameter $\eta$ that
best fits an exponential growth $I (t)\sim\exp(\eta t)$ in the
deterministic regime. For the epidemic spreading in Liberia, we find
that fitting with an exponential function on the number of cases from
July 21st to August 15th, $\eta=0.053\pm 0.003$. To find the
relation between $\eta$ and the transmission parameters $\beta_I$,
$\beta_H$, and $\beta_F$, we write the equation of the eigenvalue of
the Jacobian $J$ of the system of Eqs.~(1)--(10) for the $N_{co}=15$
counties
\begin{eqnarray}\label{eq.Jacobb}
det\Big(J(\beta_I,\beta_H,\beta_F)-\eta\;Id\Big)=0,
\end{eqnarray}
where $det$ is the determinant function, $\eta$ is the eigenvalue
obtained from the fitting, $Id$ is the identity matrix with
$10N_{co}$ rows and columns, where the factor $10$ corresponds to the
number of evolution equations for each county. Note that $J$ is a
function of the transmission coefficients.

\begin{figure}[H]
\centering
\includegraphics[scale=0.35]{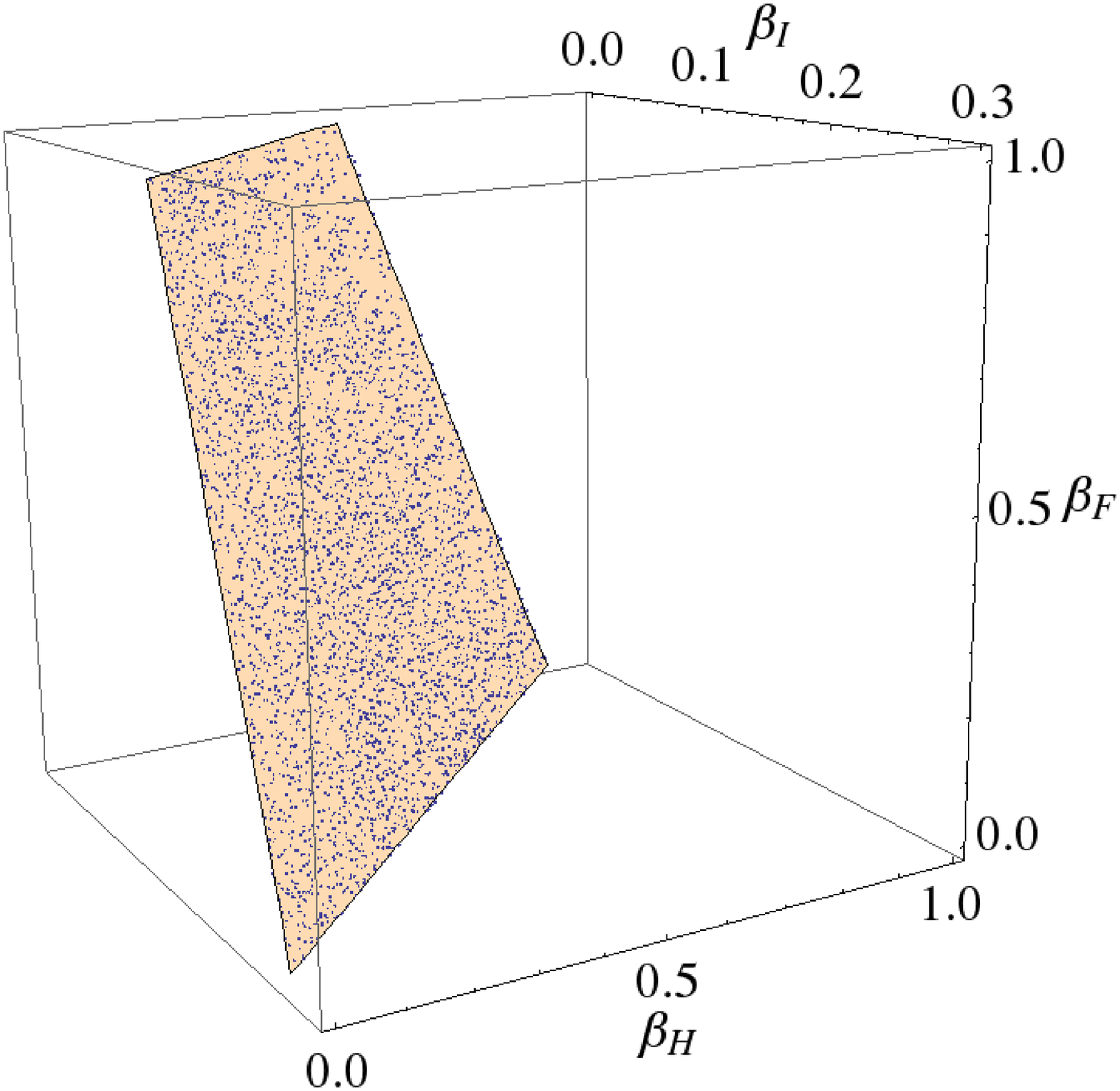}
\bigskip
\caption{Values of $(\beta_I,\beta_H,\beta_F)$ that fits the data
  obtained from i) the shift and least square method (points) and ii)
  fitting the exponential growth of number of cases (plane).
\label{f.3D}}
\end{figure}

Equation~(\ref{eq.Jacobb}) sets the relationship between the exponential
growth rate ($\eta$) and the transmission parameters, which is not
linear when the mobility is taken into account. However, we can
approximate this equation by neglecting the flow of individuals. This is
the case because although mobility spreads the EVD throughout the
country in the stochastic stage, when the disease reaches the
deterministic regime its spreading is primarily due to infected
individuals within each county and imported cases are no longer a
relevant factor. With this approximation, and using $\eta=0.054$, we
obtain an equation of a plane (see Fig.~\ref{f.3D}) that coincides with
the triads of transmission coefficients that were obtained by using the
least square fitting with a shift $s_c$. Thus this method and the
exponential fitting generate the same set of transmission parameters
that reproduces the epidemic growth.

Note that in the main text we use $s_c=200$ to calibrate the data. On
the other hand, if we use $s_c=100$ (near the 2 July date) we obtain
$\beta_I=0.09\;(0,0.23)$, $\beta_H=0.31\;(0,0.78)$, and
$\beta_F=0.46\;(0,0.99)$, and if we use $s_c=300$ (near the 28 July
date) we obtain $\beta_I=0.11\;(0,0.27)$, $\beta_H=0.38\;(0,0.94)$ and
$\beta_F=0.46\;(0,0.99)$.

\subsection*{Table of the transitions}

\begin{table}[H]
\centering
\begin{tabular}{|c|c|}
  \hline
  Transition   & Transition rate ($\lambda_i$) \\\hline
  $(S,E) \rightarrow (S-1,E+1) $ & $\frac{1}{N} (\beta_F \;S \;I
  + \beta_H \;S \;H + \beta_F\; S\; F)$ \\ 
  $(E,I_{DH}) \rightarrow (E-1,I_{DH}+1)$ & $\alpha \;\theta \;\delta E$ \\
  $(E,I_{DNH}) \rightarrow (E-1,I_{DNH}+1)$ & $\alpha \; (1-\theta) \;\delta E$ \\
  $(E,I_{RH}) \rightarrow (E-1,I_{RH}+1)$ & $\alpha \;\theta \;(1-\delta) \;E$ \\
  $(E,I_{RNH}) \rightarrow (E-1,I_{RNH}+1)$ & $\alpha \;(1-\theta) \;(1-\delta)\; E$ \\
  $(I_{DH},H_D) \rightarrow (I_{DH}-1,H_d+1)$ & $\gamma_H \;I_{DH}$ \\
  $(I_{DNH},F) \rightarrow (I_{DNH}-1,F+1)$ & $\gamma_D \;I_{DNH}$ \\
  $(I_{RH},H_R) \rightarrow (I_{RH}-1,H_R+1)$ & $\gamma_H \;I_{RH}$\\
  $(I_{RNH},R) \rightarrow (I_{RNH}-1,R+1)$ & $\gamma_I \;I_{RNH}$ \\
  $(H_D,F) \rightarrow (H_D-1,F+1)$ & $\gamma_{HD} \;H_D$ \\
  $(H_R,R) \rightarrow (H_R-1,R+1)$ & $\gamma_{HI} \;H_R$ \\
  $(F,R) \rightarrow (F-1,R+1)$ & $\gamma_{F} \;F$\\
 \hline
\end{tabular}
\caption{{\bf Table of the transition with their respective transition
    rates for our model.} Table representing the transition rates
  between different compartmental states in our model. The capital
  letters represents number of: susceptible individuals ($S$), number of
  exposed individuals ($E$), individuals infected who will be
  hospitalised and die ($I_{DH}$), individuals infected who won't be non
  hospitalised and will die ($I_{DNH}$), individuals infected who will
  be hospitalised and recovered ($I_{RH}$), individuals infected who
  won't be hospitalised and will recover ($I_{RNH}$), individuals
  hospitalised who will die ($H_{D}$), individuals hospitalised who will
  recover ($H_{R}$). Here $R$ is the number of individuals cured or dead
  and $F$ is the number of individuals in the funerals who will have
  unsafe burials and can infect. Here $\beta_I$, $\beta_H$ and $\beta_F$
  are the transmission coefficients in the community in the hospital and
  in the funerals respectively, $\delta$ is the fatality ratio and
  $\theta$ the fraction of the hospitalised ones. The inverse of the
  mean time period of the incubation is $1/ \alpha$. The mean time
  period from symptoms to hospitalisation is $1 /\gamma_H$, from
  symptoms for non hospitalised individuals to dead is $1/ \gamma_D$,
  from symptoms for hospitalised individuals to dead is $1/
  \gamma_{HD}$, from symptoms for hospitalised individuals to recovery
  is $1/ \gamma_{HI}$ and from dead to recover is $1/ \gamma_F$. The
  flow of mobility for individuals in county $ i \to j$ is explained in
  Eq.~(11)\label{table.1}.}
\end{table}
\newpage
\subsection*{Sensitivity analysis}

Because the epidemic evolution in this model is affected by many
parameters, small changes in their values could significantly impact
the model's output.  Here we analyse how changing the hospitalisation
ratio $\theta$ and the death ratio $\delta$ affects the estimation of
$\beta_I$, $\beta_H$, $\beta_F$, and $R_0$, and also affects the
evolution of the cumulative cases when the strategy is implemented in
August as described in the main text.

For $\theta$ and $\delta=0.40,\;0.50,\;0.60$ we show the values of
$\beta_I$, $\beta_H$, $\beta_F$, and $R_0$, using the least square
method and the Akaike average, as explained in the Methods section.

\begin{table}[H]
\centering
\begin{tabular}{|c|c|c|c|c|}
  \hline
  $\theta$   & $\beta_I$  & $\beta_H$  & $\beta_F$  & $R_0$\\\hline

  0.40 &0.13 &0.37 &0.40 &2.24 (1.99,2.28) \\
  0.50 & 0.14 &0.29 &0.40 & 2.11 (1.88,2.71)\\
  0.60 &0.14 &0.25 &0.39 & 2.05 (1.92,2.28) \\
 \hline
\end{tabular}
\caption{Values of the transmission coefficients and $R_0$ for
  $\theta=0.40,\;0.50,\;0.60$. Here $\delta=0.5$} \label{table.22}
\end{table}

\begin{table}[H]
\centering
\begin{tabular}{|c|c|c|c|c|}
  \hline
  $\delta$   & $\beta_I$  & $\beta_H$  & $\beta_F$  & $R_0$\\\hline
  0.40 &0.14 &0.30 &0.42 & 2.18 (1.97,2.16)\\
  0.50 & 0.14 &0.29 &0.40 & 2.11 (1.88,2.71)\\
  0.60 &0.13 &0.29 &0.38 & 2.05 (1.96,2.63)\\
 \hline
\end{tabular}
\caption{Values of the transmission coefficients and $R_0$ for
  $\delta=0.40,\;0.50,\;0.60$. Here $\theta=0.5$} \label{table.33}
\end{table}

\begin{figure}[H]
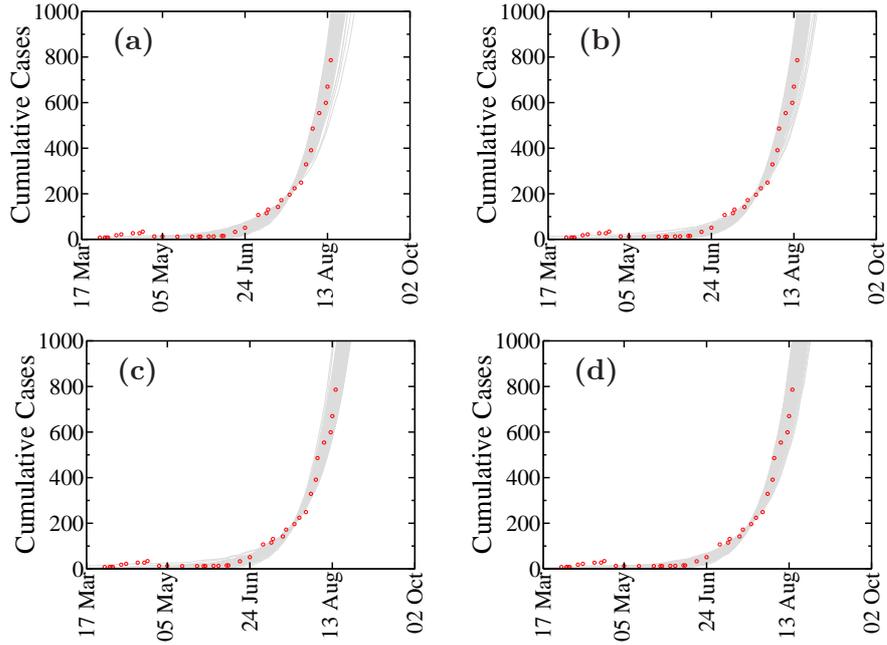

\centering
  \begin{overpic}[scale=0.2]{theta040.eps}
    \put(25,65){{\bf{(a)}}}
  \end{overpic}\vspace{0.25cm} \hspace{0.5cm}
  \begin{overpic}[scale=0.2]{theta060.eps}
    \put(25,65){{\bf{(b)}}}
  \end{overpic}\\
  \begin{overpic}[scale=0.2]{delta040.eps}
    \put(25,65){{\bf{(c)}}}
  \end{overpic}\vspace{0.25cm}\hspace{0.5cm}
  \begin{overpic}[scale=0.2]{delta060.eps}
    \put(25,65){{\bf{(d)}}}
  \end{overpic}
\caption{Evolution of the cumulative number of cases in Liberia with
  100 realisations (gray lines) and the data (symbols) with a temporal
  shift using $s_c=200$. Figures (a) and (b) correspond to
  $\theta=0.40$ and $\theta=0.60$ respectively, with $\delta=0.5$. The
  transmission coefficients used, were obtained from
  table~\ref{table.22}. Figures (c) and (d) correspond to $\delta=0.40$
  and $\delta=0.60$, respectively, with $\theta=0.50$. The
  transmission coefficients used, were obtained from
  table~\ref{table.33}.
\label{f.evolThetaDeltaCasos}}
\end{figure}

Table~\ref{table.22} shows that as $\theta$ increases 50\% from
$\theta=0.40$ to $\theta=0.60$, $\beta_H$ decreases 30\%. In contrast,
$\beta_I$ and $\beta_F$ remain almost constant and the reproductive
number does not change significantly.  On the other hand,
Table~\ref{table.33} shows that when $\delta$ increases from
$\delta=0.40$ to $\delta=0.60$, the transmission coefficients and
$R_0$ change less than 10\%, however it remains inside the
interval. Thus this model is more sensitive to changes in $\theta$
than in $\delta$.  Figure~\ref{f.evolThetaDeltaCasos} plots the number
of cumulative cases as a function of time obtained from the stochastic
model, compares the results with WHO data, and shows good agreement
between the simulations and the real data.

\bigskip

\begin{figure}[H]
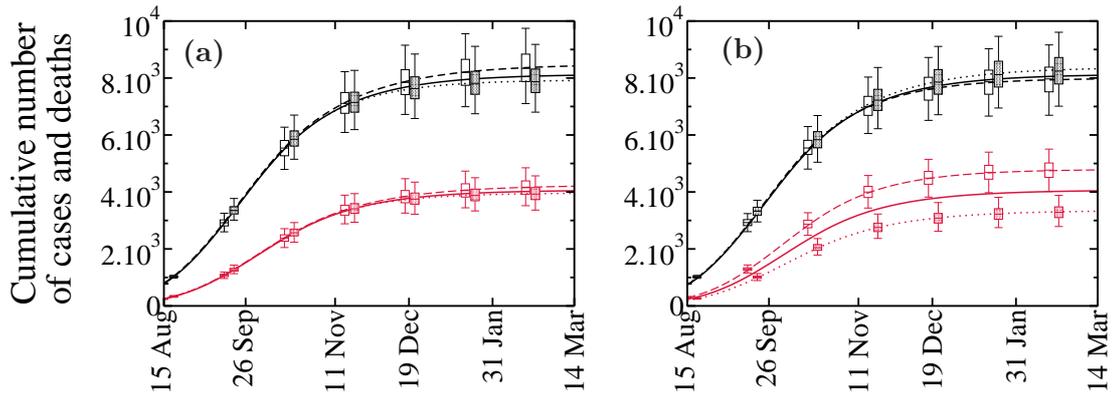

\centering
  \begin{overpic}[scale=0.25]{theta.eps}
    \put(30,58.5){{\bf{(a)}}}
  \end{overpic}\hspace{0.5cm}\vspace{0.25cm}
  \begin{overpic}[scale=0.25]{delta.eps}
    \put(20,70){{\bf{(b)}}}
  \end{overpic}\hspace{0.5cm}\vspace{0.25cm}
\caption{Evolution of the cumulative number of cases (black) and
  deaths (red) when it is applied the strategy from August, as it was
  explained in the main text, for different values of $\theta$ and
  $\delta$. In the figure (a), $\delta=0.50$ and $\theta=0.50$ (solid
  line), $\theta=0.40$ (dotted line and filled box plots) and
  $\theta=0.60$ (dashed line and open box plots). In figure (b)
  $\theta=0.50$ and $\delta=0.50$ (solid line), $\delta=0.40$ (dotted
  line and filled box plots) and $\theta=0.60$ (dashed line and open
  box plots). All the curves were obtained by integrating the evolution
  equations (1)-(10), and the box plots were obtained from the
  stochastic simulations.
\label{f.evolThetaDelta}}
\end{figure}

To evaluate how changing the parameter values alters the effectiveness
of the intervention strategy as explained in the main text,
Fig.~\ref{f.evolThetaDelta} plots the cumulative number of infected
individuals when the strategy is implemented in mid-August for different
values of $\theta$ and $\delta$.

Figure~\ref{f.evolThetaDelta} shows that although the evolution of the
number of {\it cases\/} is not sensitive to variations in $\theta$ and
$\delta$, the evolution of number of {\it deaths\/} is sensitive to
variations in $\delta$ (see Fig.\ref{f.evolThetaDelta}b). This is the
case because the final number of deaths is proportional to $\delta$,
and in our strategy $\beta_F$ changes but $\delta$ remains constant.

\noindent{\it Acknowledgements}

\noindent HES thanks the NSF (grants CMMI 1125290 and CHE-1213217) and
the Keck Foundation for financial support. LDV and LAB wish to thank
to UNMdP and FONCyT (Pict 0429/2013) for financial support.

\bigskip
\noindent{\it Contributions}

LDV, HHAR, HES and LAB designed the research, analyzed
data, discussed results, and contributed to writing the
manuscript. LDV implemented and performed numerical experiments and
simulations.

\bigskip
\noindent{\it Additional information}

Competing financial interests: The authors declare no competing financial interests.

\end{document}